# Predicting Failure: Acoustic Emission of Berlinite under Compression

Guillaume F Nataf [1,4], Pedro O Castillo-Villa [1], Pathikumar Sellappan [2], Waltraud M Kriven [2], Eduard Vives [1], Antoni Planes [1], and Ekhard K H Salje [3]

[1] Department d'Estructura i Constituents de la Matèria, Facultat de Física, Universitat de Barcelona, Martí i Franquès 1, 08028 Barcelona, Catalonia
[2] Department of Materials Science and Engineering, University of Illinois at Urbana-Champaign, IL 61801 USA
[3] Department of Earth Sciences, University of Cambridge, Downing Street, Cambridge CB2 3EQ, U K.
[4] INP Grenoble, 38031 Grenoble Cédex 1, France

**Abstract.** Acoustic emission has been measured and statistical characteristics have been analyzed during the stress-induced collapse of porous berlinite, $AlPO_4$, containing up to 50 vol% porosity. Stress collapse occurs in a series of individual events (avalanches), and each avalanche leads to a jerk in sample compression with corresponding acoustic emission (AE) signals. The distribution of AE avalanche energies can be approximately described by a power law $p(E)dE = E^{-\varepsilon} dE$ ($\varepsilon \sim 1.8$) over a large stress interval. We observed several collapse mechanisms whereby less porous minerals show the superposition of independent jerks, which were not related to the major collapse at the failure stress. In highly porous berlinite (40% and 50%) an increase of the energy emission occurred near the failure point. In contrast, the less porous samples did not show such an increase in energy emission. Instead, in the near vicinity of the main failure point they showed a reduction in the energy exponent to $\sim$ 1.4, which is consistent with the value reported for compressed porous systems displaying critical behavior. This indicated that a critical avalanche regime with a lack of precursor events occurs. In this case, all preceding large events were 'false alarms' and unrelated to the main failure event. Our results identify a method to use pico-seismicity detection of foreshocks to warn of mine collapse before the main failure collapse occurs, which can be applied for highly porous materials only.

# 1. Introduction

Porous materials are omnipresent in nature and amongst man-made materials. In mining most shafts are surrounded by porous minerals such as coal, goethite and bauxite [1,2,3]. Soils, aquifers, petroleum reservoirs, zeolites in filters, biological tissues, bones, wood, cork and man-made materials such as cements and ceramics are porous [4,5]. Under stress, the porous material which has a complex geometry with void sizes that range from nm to cm exhibit catastrophic failure due to a high concentration of collapsing voids [6]. The problem has, therefore, a multiscale, nature which has made it very difficult to formulate quantitative mechanical models [7,8].

The conditions and mechanism by which collapses occur in natural and artificial structures are widespread; e.g., fracking, implosions of mine shafts, collapse of buildings, fracture of bones, etc. [9]. Understanding the failure of porous materials under compression has important implications for their stress resistance and also for their applicability as materials and devices. Even earthquakes can be considered as a result of failures of the Earth's crust under compressive stresses and the mechanisms of the collapse bears similarities with the collapse of porous materials as a suitable model system [10]. The collapse of a structure under stress is usually not a single event, but is distributed over a multitude of smaller events which conspire to fracture the sample. The partial collapse events are named "jerks" and, within the context of out-of-equilibrium dynamics, have been classified as avalanche phenomena. Their statistical features are similar to crackling noise as reviewed by Sethna et al. [7]. Jerks are also observed under shear deformations where the microstructure of the sample changes in sudden movements rather than continuously [11,12]. Salje et al. [13] and Baró et al. [10] used a porous glass material (Vycor) to show that avalanches under compression follow almost perfect power law statistics ("crackling noise") with a characteristic critical exponent similar to those measured in mechanical instabilities in martensites and ferroelastic materials [14-17], critical dynamics in micro fracturing [18], crack growth in heterogenous materials [19], and spontaneous acoustic emission in volcanic rocks [20].These results have put the problem of understanding the failure of porous materials under compression firmly within the scenario of crackling noise and avalanche criticality.

The key question in the analysis of compression avalanches in porous materials relates to the existence of precursor effects: is it possible to predict a main event from pre-shocks before the failure event occurs? At first glance one may assume that this problem is similar to the prediction of earthquake from pre-shocks where few experimental observations exist in literature. First observations of foreshock sequences go back to 1988 were a full sequence was observed at the Chalfat earth-quake by Smith and Priestly [21] and large sequences of Californian earthquakes by Dodge et al. [22]. In each case the statistical evidence was rather limited which related to technical issues of seismological observations. Many large data sets can be obtained from laboratory experiments such as from observations in porous goethite, FeO(OH), where two scenarios were identified. Samples with porosity < 60% showed no evidence for any precursor effects and no 'early warning' signal could be extracted from the compression noise. Samples with porosities > 60%, on the other hand, did show some precursor noise and opened the possibility to use pico-seismic observations to predict the collapse of a mine [3]. This first study could not identify the physical mechanism which changed the collapsing behavior in weak and strong porous materials, mainly because very few well characterized natural goethite samples were available to explore the noise statistics in more detail. We overcame this problem by using a synthetic porous AlPO$_4$, berlinite, which could be produced in large quantities and also allowed us to explore the collapse mechanism in much more detail than previously possible.

It is the purpose of this analysis to show that the first observations in goethite were indeed correct and that the key for the change of collapse mechanism is related to an approach to critical behavior of the noise pattern near the failure stress. Similar expectations were derived previously from computer-modeling by Girard et al. [23] on the compressive failure of heterogeneous materials using a continuous progressive-damage model. However, we found increased activities only in samples with high porosity, where the typical power law statistics of the avalanches is much less well realized, than in denser materials where no such increased activities occur. Girard et al. [23] argued that the size distribution of damaged clusters lead to a critical interpretation of fracture with highly increased activity near the failure point. Friedman et al. [24] studied the compression of metallic nano-pillars and concluded that the plastic regime followed the behavior of tuned criticality. These two examples show that the observation and understanding of precursor effects near failure points is experimentally extremely demanding and theoretically controversial. It is the purpose of this article to present experimental evidence to show that precursor effects exist in porous materials but only for high porosities.

Our experimental technique for this study is the detection of the acoustic emission (AE) associated with the structural collapse. This experimental technique already revealed strong statistical similarities between the compression of natural rocks with low porosities and earthquakes over a huge interval of energies of the emitted jerks (see Refs. 10 and 25 and references there in). In porous Vycor, the distribution of event energies was shown to follow a Gutenberg–Richter behavior, with no characteristic length or time scales. The probability of a jerk with energy E follows a power law $P(E)dE \cong E^{-\varepsilon} dE$ with $\varepsilon = 1.40 \pm 0.05$. The energy interval for the power-law behavior in this experiment [10,13] spans over more than eight decades. Similar experiments in highly porous goethite showed an increase in the energy exponent from $\varepsilon = 1.68$ to $\varepsilon = 2.0 \pm 0.1$ with increasing porosity. Collapse in the hardest material analyzed so far, alumina ($Al_2O_3$), equally showed, to a good approximation, a power law distributions for the AE and the measured shape changes [9] with $\varepsilon \approx 1.8$.

**2. Sample preparation**

*2.1. Powder synthesis*

AlPO4 powder used in this study was synthesized through a simple, solution-polymerization route as reported previously [26-29]. Aluminum nitrate nonahydrate, $Al(NO_3)_3 \cdot 9H_2O$ (Alfa Aesar, Ward Hill, MA, USA) and ammonium phosphate dibasic, $(NH_4)_2HPO_4$ (Fisher Scientific, Hampton, NH, USA) were the cation sources for $AlPO_4$. Stoichiometric amounts of precursors were dissolved in deionized water and stirred for 1 h before the addition of the polymeric solution. A 5 wt% solution was made by dissolving 80% hydrolyzed PVA (Sigma Aldrich, St. Louis, MO, USA) in deionized water by stirring for 24 h at room temperature. The ratio of PVA to cation salts in the solution was adjusted in such a way that there were four times more positively charged valences from the cations than negatively charged functional end groups of the organics (in the case of PVA, −OH groups). There were more cations in solution than the hydroxyl functional groups of the polymer with which they could chemically bond. In $AlPO_4$ the total positively charged valences were 8. Since each PVA monomer has one (OH) functional group, 2 PVA monomers were used per each mole of $AlPO_4$ resulting in a cation valence to anion functional group of 4:1. Gelation did not occur during the reaction due to the acidic nature of the solution, resulting from the large amounts of nitrates and also addition of a few drops of nitric acid ($HNO_3$), which were added to ensure complete dissolution of the salts.

The precursor solutions were then heated on a hot plate with continuous stirring until the water of the solution evaporated, and a crisp, light-brown, aerated gel formed. The temperature of the solution on the hot plate varied during the drying process; however, it was generally < 300 °C. The dried gel was then ground using an alumina mortar and pestle, after which it was calcined in air at 800 °C for 1 h. For the first case, calcined powders were then attrition milled for 1 h using yttria stabilized zirconia beads with propanol as milling media in order to reduce the particle size and to increase the specific surface area, dried using a hot plate with continuous stirring to remove the ethanol, then dried at ~ 100 °C for 24 h and finally stored. For the second case, dried gels were heat treated at 1250 °C for 2 h and the particle size was reduced as explained for the calcined powders.

*2.2 Bulk sample preparation*

In order to achieve differences in the volume of the porosity, four different mixtures were attempted. (i) Bulk samples made with as-calcined and milled powders, (ii) bulk samples made with as-crystallized and milled powders, (iii) bulk samples made by adding 30 and (iv) 50 vol% graphite particles (Aldrich Chemical Company) to as-calcined powders so as to alter the porosity contents. Graphite particles which were used as pore formers in the latter case had a 1.9 g/cc density and a 1-2 μm particle size. In all these cases, 2 wt% of polyethylene glycol (PEG, Mn = 200, Sigma Aldrich, St. Louis, MO, USA) was added to the powders and ball milled with ethanol and yttria stabilized zirconia as milling media (at 100 rpm for 24 h). The dried, crushed and classified powders were initially compacted using a 19 mm cylindrical hardened steel die, uniaxially pressed under < 5 MPa and cold isostatically pressed (CIP) under 50,000 psi for 10 minutes (~ 344 MPa). A cold isostatic press (CIP, Model CP 360, American Isostatic press, Columbus, OH, USA) was employed to consolidate the powder particles homogeneously and the high pressure involved also allowed removal of intergranular pores which would result in inhomogeneity in the bulk samples during sintering. CIPed samples were then heated slowly to 900 °C at a heating rate of 2.5 °C/min to remove the pore formers (graphite particles) and binders and then heated to 1600 °C at a heating rate of 5 °C/min before being held at final temperature for 5 h in air.

*2.3 Sample characterization*

The phase formation and precursor-to-ceramic powder conversion were studied via differential scanning calorimetry and thermogravimetric analyses (DSC/TGA). Powder X-ray diffractometry (XRD) with Cu K$_\alpha$ radiation (Siemens-Bruker D5000) was used to analyze the phases present in the materials. Density of the sintered specimens was measured by the Archimedes' method in distilled water at controlled temperature (ASTM C373). Samples were sectioned with a low speed, diamond-tipped saw and cross-sectioned regions were polished by a "Buehler Ecomet III" polishing apparatus, using diamond polishing discs and polishing pads (Buehler) down to 0.25 μm. Scanning electron microscopy (SEM, JEOL 6060LV) was used to carry out the microstructural analysis on the polished surface of the specimens. A few nanometers thick layer of gold-palladium was coated on the sample surface prior to observation to ensure electronic conductivity. The sectioned surfaces from sintered samples were examined both for microstructural analysis as well as for XRD analysis.

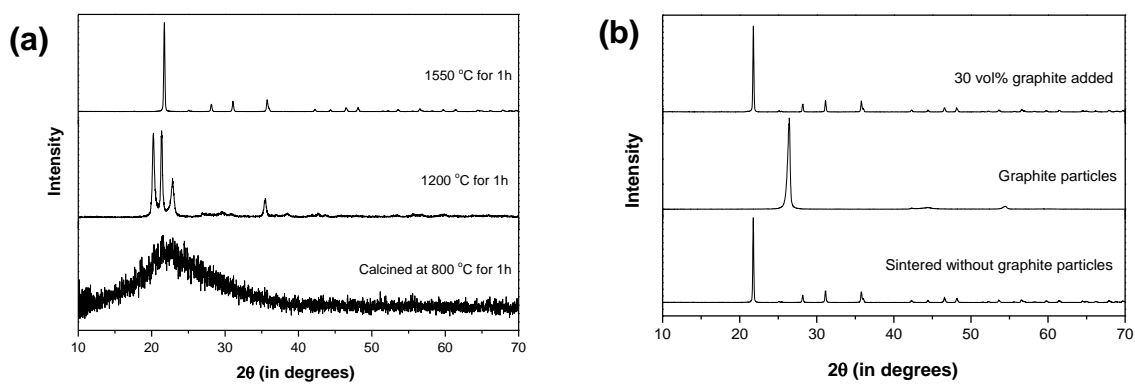

Fig. 1 Phase evaluation using XRD patterns (a) as synthesized AlPO$_4$ powders prepared by the organic/inorganic entrapment method, (b) AlPO$_4$ pellets synthesized from as-calcined powders containing 0 and 30 vol% graphite particles. The AlPO$_4$ pellets were sintered at 1600 °C for 5 h, with and without graphite particle additions, and showed only the tetragonal α-cristobalite structure.

DSC/TGA results showed that there was no weight loss observed above 800 °C in the as-prepared powders and the powders completely lost all of the organic materials well before 800 °C. Powder X-ray analysis (see Fig 1) indicated that the AlPO$_4$ powders prepared by organic/inorganic steric entrapment (the PVA synthesis route) remained X-ray amorphous even after calcination at 800 °C for 1 h. The powders started to crystallize after 1100 °C as shown in the XRD pattern of the as-prepared powder. Room temperature analyses showed that the powder samples heat treated at 1200 °C for 1 h exhibited the high temperature cubic, tridymite symmetry and the samples heat treated at 1550 °C for 1 h had the low temperature tetragonal, α-cristobalite structure. This is consistent with the operation of a critical particle size effect, whereby the nucleation of the low temperature phase on cooling is not permitted until a critical particle size has been exceeded [30]. SEM micrographs of sintered $AlPO_4$ made from as-calcined powders are seen in Fig. 2. In a microstructure of sintered samples with uniformly distributed fine porosity, pore sizes were < 10 μm. The microstructure clearly revealed that the sintered samples could not be densified under the experimental conditions that were followed, without the help of high pressure or the use of a sintering aid.

The magnified microstructure in Fig. 2 (b) reveals that the grains in the dense region of $AlPO_4$ had grown to more than 10 μm in size. Porous $AlPO_4$ samples resulting from the addition of graphite particles in the $AlPO_4$ matrix are shown in Fig. 3. The pore shapes and sizes were different than those in $AlPO_4$ sintered without any graphite particles. However, Fig. 2 indicate that the volume fraction of pores in the $AlPO_4$ sintered without graphite particle was less compared to the graphite added sample (Fig. 3). When amorphous powders undergo sintering, in addition to the regular diffusion mechanism, the re-arrangement due to crystallization (volume change) may also induce microcracks and hence porosity. Bulk density and apparent porosity values measured by the Archimedes technique are given in Table 1. Preliminary density and porosity analyses showed that the porosity values changed with changing the starting $AlPO_4$ powders and also by increasing the volume of graphite particles in the $AlPO_4$-graphite mixture. Apparent porosity values were calculated using the measured bulk density values and the theoretical density value of $AlPO_4$ (2.566 g/cm$^3$).

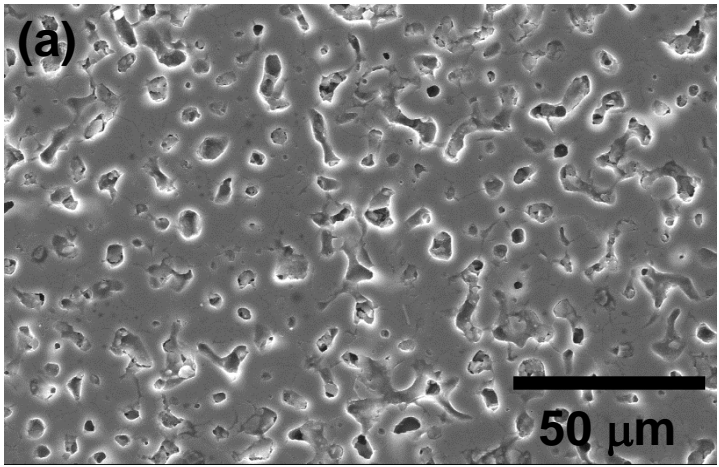 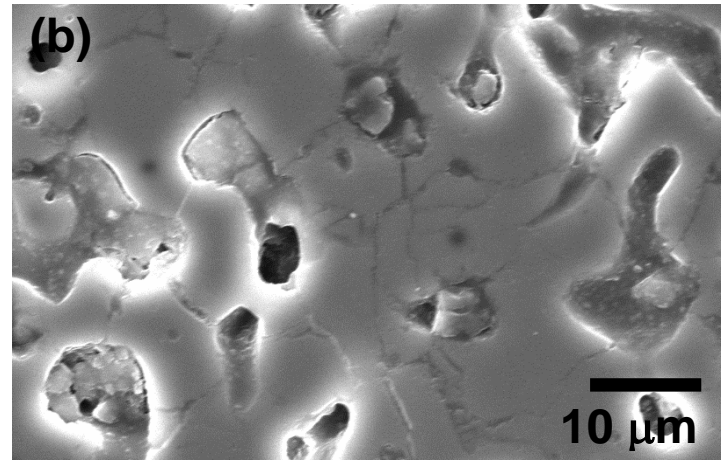

Figure 2. SEM micrographs of polished AlPO$_4$ samples sintered at 1600 ºC for 5 h using as-calcined powders (without any added graphite particles).

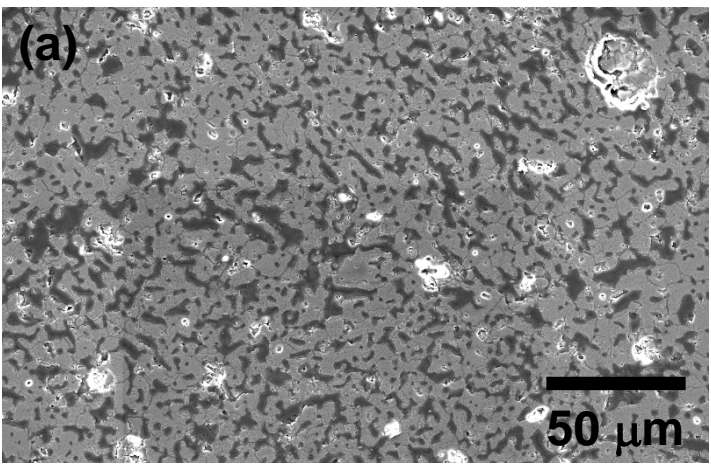 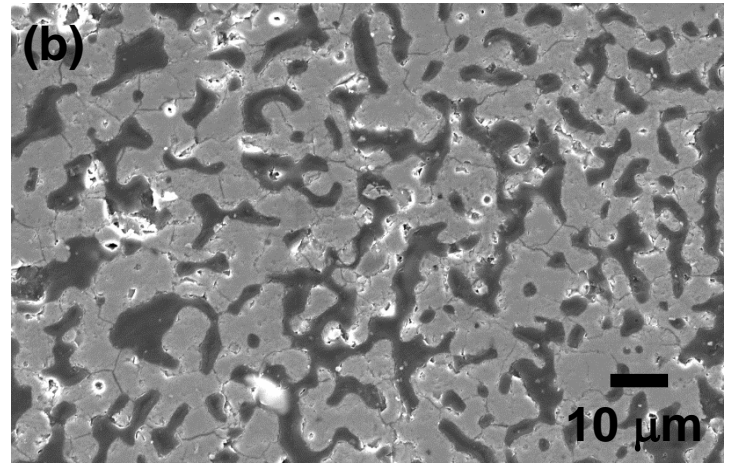

Fig.3 SEM micrographs of polished AlPO$_4$ samples sintered at 1600 º C for 5 hours using as-calcined powders (with 30 vol% added graphite particles).

Table 1 Summary of dimensions, apparent porosity, compression conditions, and measured compressive strength for the specimens studied. The sections of the specimen were determined as the surface area limited by their external perimeter.

| Sample ID | Section (mm$^2$) | Mass (mg) | Apparent porosity (%) | Stress rate (kPa/s) | Compressive strength (MPa) |
|---|---|---|---|---|---|
| AlPO$_4$-50 | 74.8 | 70.2 | 50.7 +/- 0.03 | 3.1 | 1.9 |
| AlPO$_4$-30 | 44.2 | 53.3 | 39.1 +/- 0.05 | 5.3 | 9.6 |
| AlPO$_4$-calc | 43.6 | 52.1 | 32.3 +/- 0.05 | 2.7 | 15.4 |
| AlPO$_4$-cry | 44.1 | 40.2 | 30.0 +/- 0.10 | 3.0 | 22.2 |

## 3. Acoustic emission measurement

The experimental arrangement for the uniaxial compression setup has been described elsewhere [9, 10,13]. It consisted of two parallel circular aluminum plates perpendicular to the vertical direction (see inset of Fig.4). The bottom plate, hanging from the load cell at the top of the arrangement, was static. The upper plate was pulled downwards by means of three guides sliding through precision ball bearing elements mounted on convenient holes drilled in the bottom plate. The pulling device consisted of a water container acting as a dead load. Small pump rates for the inflowing water enabled the imposition of a slowly increasing load. An acoustic emission sensor was embedded into the upper compression plate, as depicted in the inset of Figure 4. The sensor used was a model micro-80 from Physical Acoustics Corporation and it was placed 4 mm away from the specimens. It was encapsulated in stainless steel in order to reduce electrical noise and it had a broad band frequency response extending from ~175 kHz to ~1Mhz (maximum sensitivity of 0.3 V/mbar). A thin vaseline layer was used between the compression plate and the sensor and between the sample and the compression plate, in order to ensure a good ultrassound acoustic coupling. The signal from the sensor was pre-amplified to 60 dB and input in a PCI-2 system (Europhysical Acoustics, Mistras group, France) operating at 10 MHz and with a digital pass band filter 1 kHz-2 MHz. A laser extensometer (Fiedler Optoelektronik, Germany) measured the vertical separation z between the plates to a resolution of 100 nm. The load cell (1 kN range) signal was read with a lock-in amplifier and had been calibrated with standard weights. Possible noise arising from the friction of the guides with the bottom compression plate was first calibrated using blank experiments. A software filter was then employed to all measurements in order to identify and suppress signals originated from this source.

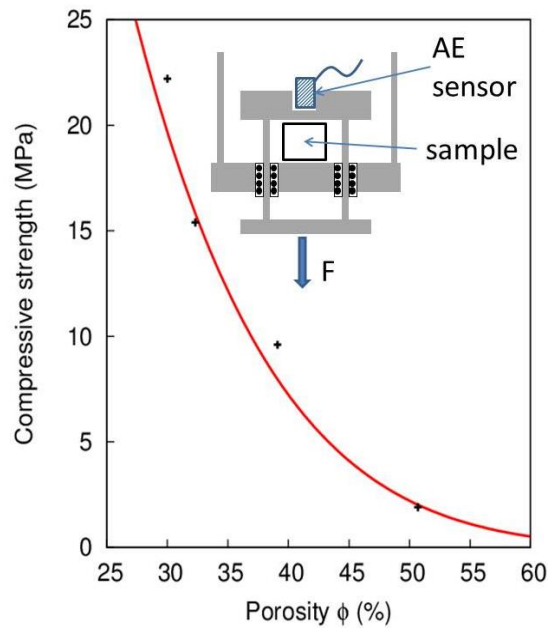

Fig. 4 Compressive strength of berlinite versus porosity. The line is the interpolated power law Pc= 200 (1-Φ/100) $^{6.5}$ [6]. The inset is a schematic representation of the compression device.

We performed an avalanche analysis from the acoustic emission signal. The beginning of an avalanche event (hit) was defined as the time $t_1$ at which the voltage from the transducer exceeded a predefined threshold (27dB). The end of the event $t_2$ occurred when the voltage remained below threshold for more than 100 μs. The energy E of every event was computed as the integral of the square voltage between $t_1$ and $t_2$, normalized by a reference resistance. The macroscopic compression process has been monitored with the two averaged quantities: (i) The acoustic activity, measured as the number of hits in time units of ten seconds and (ii) the energy emission, measured as the sum of energies of the individual hits recorded every 10 seconds.

**4. Results**

Results presented in this section were obtained from the specimens listed in Table 1. We have checked that similar results could be obtained with samples of same porosity and comparable geometrical characteristics cut from the same ingots. We have also checked that results were little affected by small changes in the position and acoustic coupling of the transducer and studied sample in the compression device. Hence, the location of the transducers and samples were kept identical for all the experiments performed in this study.

The failure stresses of four samples and a theoretical interpolation based on a statistical model by Salje et al. [6] are shown in Fig. 4. In this study, stresses were simply estimated as the ratio of the applied force and the section of the studied specimen given in Table 1. In a previous study of porous alumina under compression [9], compressive strengths between 25 and 250 MPa were fitted with the behavior $P_c = 70 (1-\Phi/100)^m$ with m = 3.8. In the present case, the same relation seems to apply for smaller compressive strengths ($P_c < 25$) but with a higher exponent m = 6.5. Examples of the AE recordings for four samples are shown in Fig 5. The failure of each sample is clearly seen in the upper panel, showing the height z versus time. The collapse of the sample occurs at the smallest stress for the highest porosity (50%) and increases with decreasing porosity to the samples with 30% porosity ($AlPO_4$-cry). The rapid sample contraction is measured by $J(t) = (dz/dt)^2$ where z is the height of the sample and t is the time elapsed under constant stress rate (and is hence proportional to the applied stress). The jerk function (second panels from top) J(t) represents a first indication of the jerk distribution and are compared with the acoustic activity (third panels) and the energy emission (bottom panels). Several conclusions can be reached simply from visual inspection of Fig. 5. Along with the compression process, both the acoustic activity and the emission energy exhibit a rather variable behavior, characterized by several peaks that precede the main failure. Even after the failure event some further activity occurs in the debris. For the high density samples (denoted as calc and crys) many large events took place before the failure so that a detailed 'early warning' system cannot be envisaged. The number of false alarms was high and randomly distributed. On the contrary, for the most porous samples, the first clear increase in the energy emission and AE activity could be clearly correlated with the main failure.

*4.1 Statistics of AE events*

The distribution of individual AE jerks over the full compression process was considered for all the studied specimen. Fig. 6 shows the probability function to find a jerk event P(E) with energy in the interval between E and E+dE. The curves followed the same power-law behavior, $p(E) \sim E^{-\varepsilon}$, over 5 decades with few subtle differences. The power-law exponent indicated by the straight line corresponds to the value $\varepsilon=1.8$ in agreement with the value reported for goethite [3] and alumina [9].

The power-law exponents could be obtained from the data using the maximum likelihood method (see Ref. 31). The exponent was fitted by considering a higher cutoff equal to the maximum energy measured, and a lower cutoff varying within several orders of magnitude. It is expected that it should exhibit a plateau when the fitted exponent is represented as a function of lower cut-off.

While the overall trend follows a power law with an exponent 1.8 we found a weak decay of P(E) for large energies for the most porous minerals. The activity of the acoustic emission was also statistically invariant for all the time intervals and all the samples. This leads to a first conclusion that the emission activity and the overall energy exponents are - within experimental uncertainties- the same for all the samples and hence for all the porosities. This result agrees with the analysis of micro- and pico-seismicity in boreholes by Davidsen and Kwiatek [32]. For further comparison we have equally analyzed waiting time distributions $D_{E,R}(\delta)$, where $\delta_j = t_j - t_{(j-1)}$ is the waiting time between two consecutive events with energy larger than a given threshold energy $E_{min}$ (which takes values from $10^{-1}$ aJ to $10^2$ aJ). In Fig. 7, these distributions have been plotted in a scaled representation for all the studied samples and different values of $E_{min}$ (indicated in the figure). In order to compare the shape of the distributions we have rescaled the axes as $\langle r_{E_{min}} \rangle^{-1} D_{E_{min}}(\delta)$ and $\langle r_{E_{min}} \rangle \delta$ respectively, where $\langle r_{E_{min}} \rangle$ is an average activity given by the mean number of events per unit time with energy $E$ larger than $E_{min}$. To a very good approximation the different distributions collapse into a single one, which proves the existence of a scaling law. The lines in Fig. 7 represent the slopes for the two power law regimes obtained from similar measurements on Vycor [10]. The axes have been rescaled as $\langle r_{E_{min}} \rangle^{-1} D_{E_{min}}(\delta)$ and $\langle r_{E_{min}} \rangle \delta$ respectively; where $\langle r_{E_{min}} \rangle$ is the mean number of events per unit time with energy $E$ larger than $E_{min}$. For every value of $E_{min}$, the number of analyzed hits is indicated between parentheses in the legend. The two black lines show the equivalent waiting times in Vycor with two power law distributions. The short waiting times have an exponent of -0.93, the longer waiting times correspond to a higher exponent – 2.45.

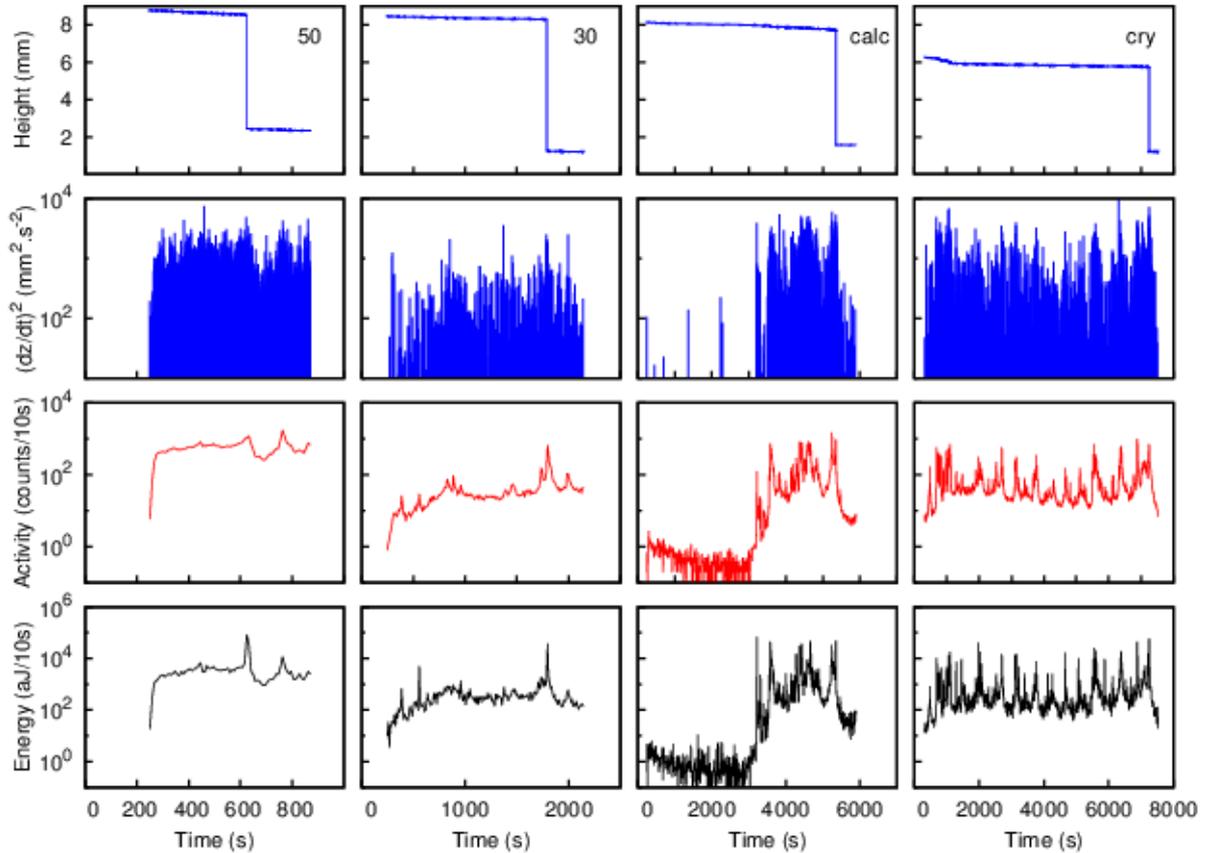

Fig. 5 Time evolution of the compression of porous berlinite (AlPO$_4$), for the four samples in Table 1 with different degrees of porosity. The upper panels show the length change of the sample where the main failure event is seen as a major sample collapse. The second set of panels shows the micro-collapses evaluated as the square of the length derivatives with time (J(t) = (dz/dt)$^2$ ) where each 'jerk' J(t) indicates the collapse of some cavities leading to a macroscopic length change. The third row of panels shows the acoustic emission activity and the lower panels show the energy emitted in the acoustic emission signal. Major precursor peaks in the emitted energy are clearly visible 5 s and 7 s before the main collapse in the samples with 39.1% (AlPO$_4$-30) and 50.7 % (AlPO$_4$-50) porosity, respectively.

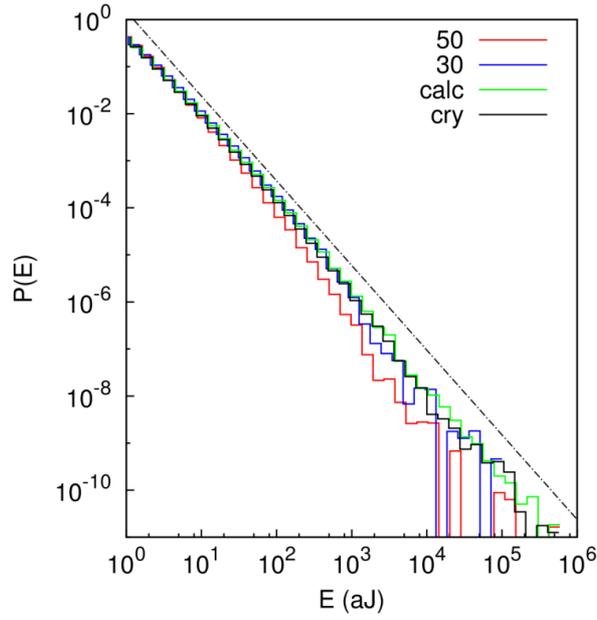

Fig. 6. Energy distribution of the compression jerks measured by acoustic emission. All four samples are super imposed. The energy exponent is drawn by the dotted line as -1.8.

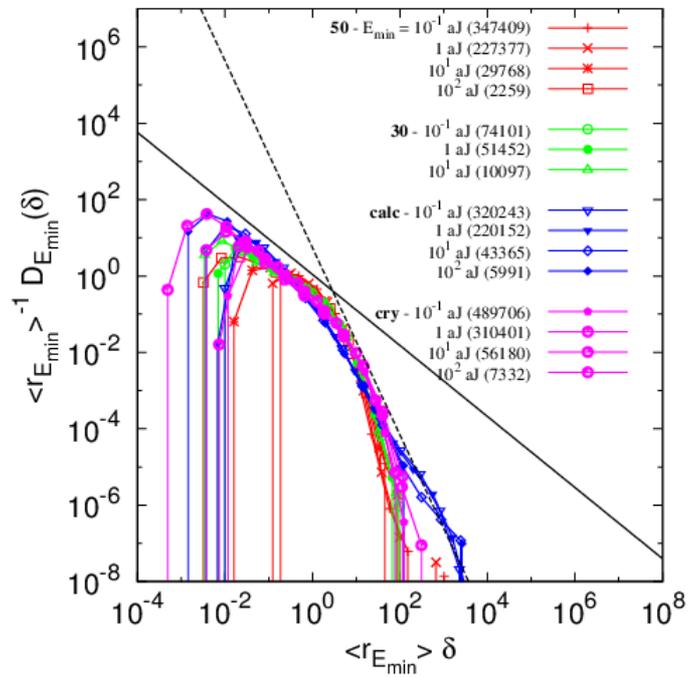

Fig.7 Scaled representation of waiting time distributions (see text) for all samples and several energy intervals corresponding to different values of $E_{min}$ which are indicated in the legend.

*4.2 Small time intervals near the big failure for high porosity samples*

In Figs. 8-10 we show the histogram P(E) for the three samples, $AlPO_4$-30, $AlPO_4$-calc and $AlPO_4$-cry, analyzed for limited time intervals close to the big failure. As observed in Fig. 5, while the observed precursor signals were similar in these three samples, we found that the largest plateau in the exponent *versus* $E_{min}$ plots corresponded to the sample with 32% porosity (Fig. 8). This sample reveals a 'classic' power law behavior with a well-defined exponent of 1.4 near the main failure event, while the power law statistics were less defined for times outside this interval (purple curve in Fig. 8). The overall exponents were larger and fit the average exponents of all curves (1.8). The similar sample with 30% porosity (Fig. 9) did not follow the same trend, although the effective exponents tended to become smaller near the main failure event for all samples. A weak plateau at slightly higher exponents (1.55) occurred for a sample with 39% porosity in Fig. 9.

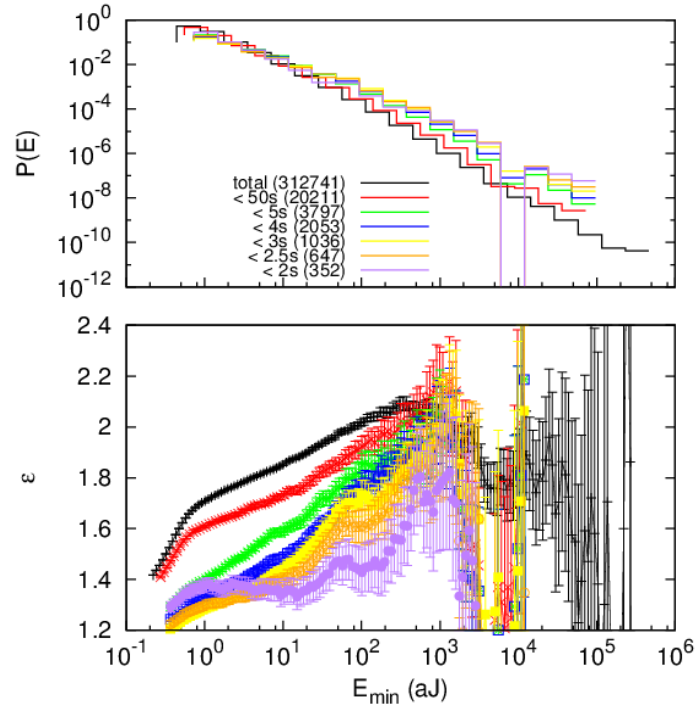

Fig.8 Probability distribution of the acoustic emission signals for a sample with 32% porosity ($AlPO_4$-calc). The lower panel shows the exponents derived from the maximum likelihood analysis. Criticality corresponds to large plateaus in this graph which were only observed for acoustic emission just before the main failure event (lower curve). The exponent in this regime is was 1.4, All other intervals show finite slopes in $\varepsilon$ ($E_{min}$) with higher effective exponents near 1.8.

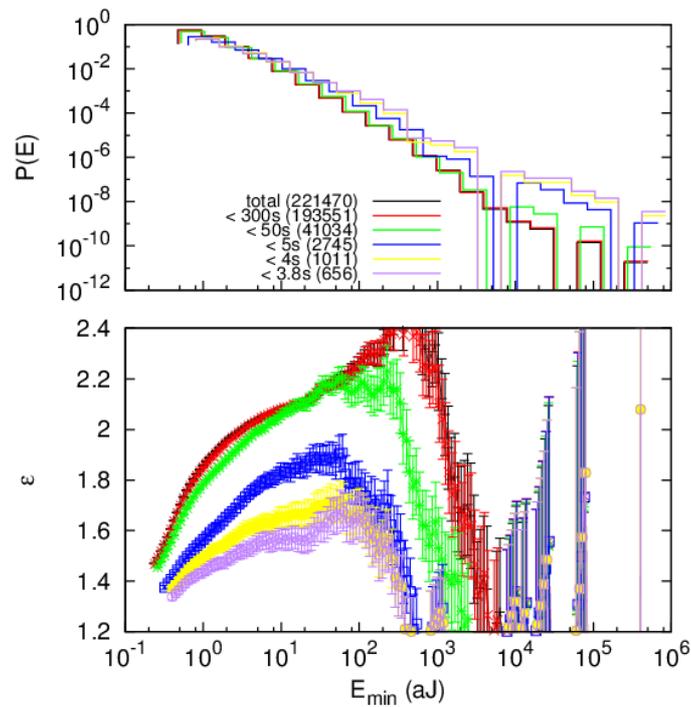

Fig. 9 Probability distributions of the acoustic emission signals for a sample with 39% apparent porosity (AlPO$_4$-30). The plateau regime becomes larger when the acoustic emission signals were restricted to the precursor regime (purple curves) with a lower effective exponent that in all other intervals.

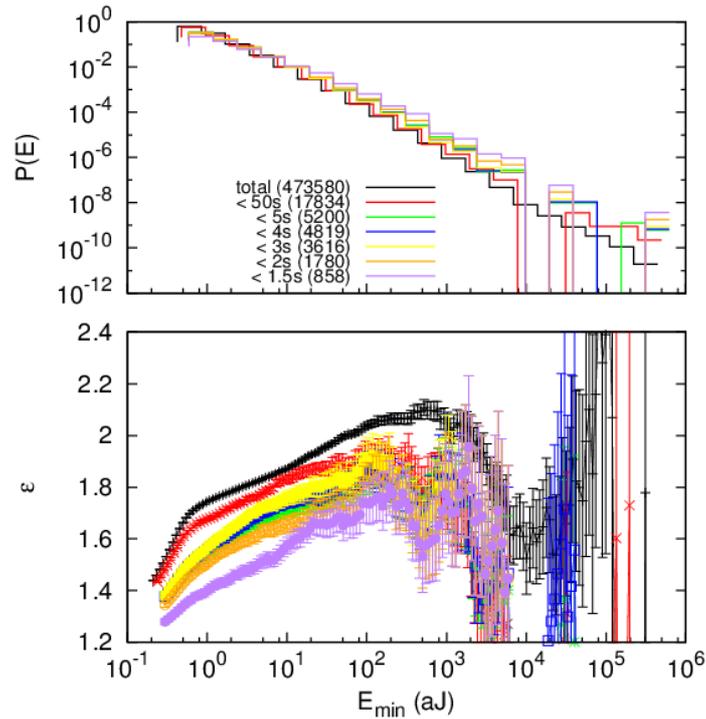

Fig. 10 Probability distributions of the acoustic emission signals for a sample with low porosity, 30 vol% (AlPO$_4$-cry). No plateau regimes develop while the effective power law exponent remains a function of the chosen time interval of the experiment. The apparent exponent seen in the power law distribution was ~ 1.8.

## 5. Discussion

We found experimental evidence for precursor effects before a major collapse event in porous berlinite (AlPO$_4$). The precursor events were found only in highly porous material but not in relatively denser samples. This observation confirms a similar tendency in a typical mining material, namely goethite [3]. Both materials exist in the geological context as porous minerals so that our observations are directly relevant to the observation of warning signs in collapsing mines. The approximate power law dependence of the jerks signals encourages us to think that the scale invariant of the observation is, at least approximately, valid, so that small and large, geological events can be described by the same power laws. Surprisingly, the collapse of $SiO_2$ based materials (Vycor) was shown previously to scale with even larger events, such as earthquakes in California [10].

The distinguishing feature of collapses in various porous materials is the time scale over which the collapse occurs. The strain rate in our experiments was ~ $3\times10^{-6}$ m/sec for the most porous berlinite, which is close to man-made stress rates in a mining scenario (typical advance of a mineshaft in coal mining). This strain rate is slower than that of breaking bones in accidents and much slower than shock absorption of porous materials under ballistic impact, say in shock absorbers in flack-jackets. It is faster by some 3 orders of magnitude than tectonic movements, which lead to earthquakes. We are undertaking computer modeling to explore the rate dependence of acoustic emission, which is unknown so far.

The second open question is related to the universality class of the porous collapse. Porous collapse under stress is not necessarily the same universality class as depinning transitions of Barkhausen noise of strongly disordered magnetic materials and plastic charge density wave depinning [33], which is yet different from the universality class of soft magnets and crystal plasticity. One possible explanation of our experimental finding is that the collapse behavior of porous Vycor (as a prominent member of this class) constitutes a universality class with power law statistics and an energy exponent near 1.4. This universality class would then be observable also in berlinite close to the major failure collapse in materials with porosities near 32%. All other regimes seem to display effective exponents near 1.8 with no significant plateau in the exponent *versus* $E_{min}$ curves. One might be tempted to allocate a different universality class to these materials. Significant precursor activity is only observed in this group of materials. An alternative consideration lies in the differences in microstructure leading to the observed differences in mechanical behavior, specifically grain sizes and porosity, as seen in Figs. 2 and 3.

The answer to this question has significant implications: it remains unclear whether earthquakes in a given geological region fall into the same universality class. If they do, then we can understand why laboratory experiments reproduce so extraordinarily well the earthquake dynamics. This would mean that we may extrapolate our findings of precursor shocks to some earthquakes (or their absence to others).

**Acknowledgements**

We acknowledge financial support from CICyT, (Spain) (Project Nº MAT2013-40590-P). E.K.H.S. is grateful for support by the Leverhulme fund (RG66640) and EPSRC (EP/K009702/1). PC-V acknowledges support from CONACYT (Mexico) under scholarship No. 186474. PS and WMK acknowledge a United States Army Research Office MURI grant (W911NF-09-1-0436), through Dr David Stepp. The scanning electron microscopy (SEM) and X-ray analysis works were carried out in the Frederick Seitz Materials Research Laboratory at the University of Illinois at Urbana-Champaign.